\documentstyle[aps,preprint,times,epsfig]{revtex}
\newcommand{\br}{{\bf r}}
\newcommand{\bx}{{\bf x}}
\newcommand{\bdr}{\delta {\bf r}}
\newcommand{\bdri}{\delta {\bf r}_i}
\newcommand{\bdx}{\delta {\bf x}}
\newcommand{\etab}{\mbox{\boldmath $\eta$}}
\newcommand{\xibi}{\mbox{\boldmath $\xi_i$}}
\newcommand{\xibj}{\mbox{\boldmath $\xi_j$}}

\newcommand{\BE}{\begin{equation}}
\newcommand{\EE}{\end{equation}}
\newcommand{\BA}{\begin{eqnarray}}
\newcommand{\EA}{\end{eqnarray}}

\begin{document}
\draft
\title{
The role of diffusion in the chaotic advection of a passive scalar with
finite lifetime
}
\author{ Crist\'obal L\'opez$^{1}$ and
Emilio Hern\'andez-Garc\'\i a $^{2}$
}
\address{
$^{1}$ Dipartimento di Fisica,
Universit\`{a} di Roma `La Sapienza', P.le A. Moro 2, I-00185,
Roma, Italy \\
$^{2}$Instituto Mediterr\'aneo de Estudios Avanzados, IMEDEA
(CSIC-Universitat de les Illes Balears), 07071 Palma de Mallorca,
Spain
}
\date{\today}
\maketitle

\begin{abstract}
We study the influence of diffusion on the scaling properties of
the first order structure function, $S_1$, of a
two-dimensional chaotically advected passive scalar with finite
lifetime, i.e., with a decaying term in its evolution equation. We
obtain an analytical expression for $S_1$
where the dependence on the diffusivity, the
decaying coefficient and the stirring due to the chaotic flow is explicitly
stated.
 We show that the presence of diffusion
introduces a crossover length-scale, the diffusion scale ($L_d$), such that
the scaling behaviour for the structure function is analytical for 
length-scales shorter than $L_d$, and shows a scaling exponent that
depends on the decaying term and the mixing of the flow for larger scales.
Therefore, the scaling exponents for scales larger
than $L_d$ are not modified with respect to those
calculated in the zero diffusion limit. Moreover, 
$L_d$ turns out to be independent of the decaying coeficient, being its
value the same as for the passive scalar with infinite lifetime.
 Numerical results support our theoretical findings.
 Our analytical and numerical calculations rest upon the Feynmann-Kac
representation of the advection-reaction-diffusion
partial differential equation.

\end{abstract}
\pacs{PACS: 47.52.+j, 05.45.-a, 47.70.Fw, 47.53.+n}


\section{Introduction}
The subject of advection of (chemically or biologically) reacting substances is 
of major importance in fields ranging from combustion to water quality control
(see for example \cite{muchosintro}).
One of the simplest models of reaction occurring  in a flow is the decrease at a constant
rate  in the
concentration of a substance. Radioactive or photochemical decay belong to this class
of reactions. The spatial structure of the decaying chemical under chaotic
fluid stirring
has been described in recent works \cite{prl,pre}
 with emphasis on the microstructure formation by the 
stretching properties of the flow, at scales at which molecular diffusion can be neglected.
The interpretation of real data, specially in laboratory experiments in which the diffusion
scale may be explicitely resolved, requires however the consideration of diffusive
processes which tend to homogeneize the structure at the smallest scales. 

 In this paper we 
investigate the role of the molecular diffusion in the 
scaling properties of 
the first order structure function, defined below
(\ref{sfq}) when $q=1$, 
for a passive scalar advected by a two-dimensional flow
yielding Lagrangian chaos, and with finite lifetime. Thus, the
passive scalar field (passive in the sense that there is no back
influence on the hydrodynamics) $\phi ({\bf x}, t)$ evolves
according to 

\begin{equation}
\frac{\partial \phi({\bf r},t)}{\partial t} + {\bf v}
({\bf r},t)\cdot \nabla  \phi({\bf r},t) =
\nu \nabla ^2 \phi({\bf r},t)-b \phi({\bf r},t)+
S({\bf r}),
\label{ard}
\end{equation}
where $\nu$ is the diffusivity, $-b\phi$ represents the decaying and
indicates that the passive scalar field decays with time at  rate
$b>0$, $S({\bf r})$ is a steady source of the passive scalar, and finally,
${\bf v} ({\bf x},t)$
is the velocity of the flow. Here we assume that the velocity
field is two-dimensional, incompressible, smooth, and nonturbulent.
Chaotic advection in two dimensions is obtained generically if a time dependence
(periodic, for e.g.)
is included in ${\bf v}({\bf x},t)$ \cite{chaadv}.

Eq. (\ref{ard}) can be considered as an approximation to more complex
biological or chemical phenomena, like 
the dynamics of plankton populations in ocean currents \cite{abraham,pce}
and
the advection of pollutants or chemical substances in the atmosphere
\cite{peter}, or as a description of simple procceses, like the relaxation
 of the sea surface temperature  \cite{abraham2}.
 Moreover,  in \cite{Keeyol} it is argued
that, in a determined scale range, the vorticity of a turbulent flow
may evolve passively 
according to Eq. (\ref{ard}), indicating the high relevance
of this equation in the studies of two-dimensional turbulence.

In the different case of a turbulent velocity field, the 
spatial propiertes, neglecting intermittency corrections,
 of a scalar field evolving through Eq. (\ref{ard})
were first written down by Corrsin \cite{corrsin}. 
The analytical expression for the
 power spectrum was obtained, among other situations, in the Batchelor's regime,
 that is, 
between the smallest typical length scale of the velocity field and the
characteristic length scale of diffusion ($L_d$),  and from this, the
scaling exponents and the diffusion length scale, $L_d$, were derived. 
The calculated form of the power spectrum $\Gamma (k)$ is
\begin{equation}
\Gamma (k)= N k^{(2b/\lambda)-1} exp(-2 \nu k^2/\lambda),
\label{corrs}
\end{equation}
where $N$ is a constant and
$\lambda$ is the absolute value of the most negative
of the average stretching rates.
 However, 
 no numerical test of this result, checking  the influence of 
molecular diffusion on  the crossover, has been
reported. Moreover, the criticisms made by Kraichnan \cite{kraichnan} to
the crossover of
the power spectrum calculated by Batchelor
\cite{batchelor} for  a passive scalar with infinite 
lifetime, $b=0$, as being too sensible to 
the intermittency, can also be applied here.

In this paper we study the case of a smooth (nonturbulent)  chaotic flow, and
instead of the power spectrum we calculate the first order structure function.
In the absence of diffusion, the scaling exponents
 for the structure functions
were calculated 
neglecting intermittency corrections  \cite{prl},
or including them for smooth chaotic flows \cite{pre,nam},
and
in \cite{chertkov}  for the Kraichnan flow in the spatially
smooth limit. 
 The structure functions are 
important  quantities than can yield information about the typical variation
of the concentration field over a small distance $\delta x$.
Thus, a structure function of order 
$q$ is defined as
\begin{equation}
S_q(\delta x)=<|\delta \phi|^q>=<|\phi({\bf x}+\delta x{\bf n})-\phi({\bf x})|^q>,
\label{sfq}
\end{equation}
where $< >$ denotes spatial average, ${\bf n}$ is a unit vector in the 
chosen direction,  and $q$ is a positive real number. In general, for small
$\delta x$ the structure functions are expected to exhibit a power-law dependence
$S_q(\delta x) \sim \delta x^{\zeta _q}$ characterized by the set of scaling exponents
$\zeta_q$.
Neglecting intermittency corrections implies that the 
 scaling exponent  $\zeta_q$
varies linearly with $q$. In this work we neglect intermittency, and
therefore, we can limit ourselves to  study only the first order structure function,
 $S_1(\delta x)$.
It is important to note that working with the first order structure function
will minimize the effect of the intermittency
\cite{vulpio} when comparing our analytical results with
numerical (as it is performed here) or real data.
 In this sense, we improve the calculations
performed by Corrsin because the intermittency corrections are more important
for the power spectrum than for the first order structure function.

Summing up, in this work we study the spatial properties of a scalar field evolving 
with Eq. (\ref{ard}) when the velocity field is  smooth and chaotic.
 We calculate the first order
structure function and check the influence of the diffusion on its scaling
exponent and crossover. 
We also perform numerical calculations to check our analytical 
results.
Numerically we show that effectively the calculations performed with
the first order structure function improve (\ref{corrs}).
Finally, we mention that we base all our analytical and numerical  calculations
 in the Feynman-Kac representation of
  Eq. (\ref{ard}) \cite{freidlin,falkovich}. In other words, we try to obtain 
information about the spatial properties of an Eulerian field $\phi (\bx,t)$ in
the long-time limit, through calculations following individual fluid trajectories,
i.e., using a Lagrangian formulation \cite{falkovich}. 
  Our numerical code is
explicitly stated in the {\it numerical results } section.

\section{Analytical calculations of the first order structure function}

The solution of Eq.~ (\ref{ard}), with initial condition
$\phi(\bx,0)=\phi_0(\bx)$ can be written in terms of the so-called
Feynmann-Kac representation, \cite{freidlin}
\begin{equation}
\phi (\bx,t)  = \left< \phi_0[\br(0)] e^{-bt} +
\int_0^t ds e^{-b(t-s)} S[\br(s)] \right>_{\etab}
\label{FK}
\end{equation}
where $\br(t)$ is the solution of the Langevin equation
\begin{equation}
\frac{d{\bf  r}}{dt}  =  {\bf v}({\bf  r},t)+
\sqrt{2\nu}\etab(t) \label{langevin}
\end{equation}
which satisfies the {\sl final} condition $\br(t)=\bx$.
It is worth to mention that (\ref{FK}) considers the backwards-in-time
dynamics and  the problem is well-posed by fixing the final conditions instead of
the inital ones. 
 $\etab(t)$
is a normalized vector-valued white noise term with zero mean,
i.e. $<\etab(t)>=0$ and $<\etab(t)\etab(t')>={\bf I}\delta(t-t')$,
with ${\bf I}$ the identity matrix. The average $\left< \cdot
\right>_{\etab}$ is taken over the different stochastic trajectories
${\bf  r}(t)$ ending at the stated final point $\bx$.
 Note that
$\etab$ is a dummy variable in (\ref{FK}), which disappears after
the averaging. In consequence, in expressions containing several
$\phi$'s, for example at different space points or times, a
different noise variable, each one statistically independent from
all the others, should be introduced for every appearance of
$\phi$.

It is important to note that in the long-time limit
 the $\phi$ field
does not approach a steady distribution
but one with the same time-dependence of the flow. Thus, for time-periodic 
flows  the $\phi (\bx,t \to \infty)$ field is also time-periodic. 
 However, its singular characteristics do not change in time,
that is, it is a statistically steady field \cite{prl,pre}. 
 In the following we focus in the situation in which the
initial concentration is $\phi_0=0$, so that all the structure
arises from the source. 
Our
analytical calculations follow closely along the steps of
\cite{prl,pre,pce}, thus we submit the reader to these references for
further details. We proceed by calculating the difference at time
$t$ for the values of the chemical field at two different points
$\bx+\bdx/2$, and $\bx-\bdx/2$ separated by a small distance
$\bdx$.
The expression for the difference $\delta\phi(\bx,t;\bdx)
\equiv \phi(\bx+\bdx/2,t)- \phi(\bx-\bdx/2,t)$
 is
\begin{equation}
\delta\phi(\bx,t;\bdx) =
 \left<  \int_0^{t} ds e^{-b (t- s)} \delta S[\br_1(s),\br_2(s)]
 \right>_{\etab_1,\etab_2}   \ ,
\label{deltaC1}
\end{equation}
where
\BE
\delta S[\br_1(s),\br_2(s)] \equiv
S[\br_1(s)]  -  S[\br_2(s)]  \ .
\label{deltaS}
\EE
Here $\br_1(s)$ and $\br_2(s)$, with $0<s<t$, are the solutions of
Eq.~(\ref{langevin}) with driving noise terms $\etab_1$ and
$\etab_2$, respectively, and final values $\br_1(t)=\bx+\bdx/2$,
and $\br_2(t)=\bx-\bdx/2$. Two independent noise processes
$\etab_1$ and $\etab_2$ have been introduced since two
concentration values are used in Eq.~(\ref{deltaC1}). We are
interested in the asymptotic ($t\rightarrow \infty$) scaling of
Eq.~(\ref{deltaC1}) for $\bdx$ finite but small. The contributions
in the integral (\ref{deltaC1}) split into very different
behaviors: there is a time $t_s$ such that if $t_s<s<t$, the
backwards trajectories $\br_1(s)$ and $\br_2(s)$ exponentially
increase its initial distance $\bdx$, since they are advected by a
chaotic flow. In this regime, the difference $\delta S$ will also
grow. At $t_s$ the trajectory separation is of the order of the
system size, or of some coherence length of the advection
velocity, $L$, and thus can not continue to grow. Thus, for
$s<t_s$, $\delta S$ will stop its systematic growing and
$S[\br_1(s)]$ and $S[\br_1(s)]$ will fluctuate between the range
of values taken by $S$ in the system, in a manner independent on
the initial separation $\bdx$. Since there is no longer difference
in the statistical properties of $\int_0^{t_s} ds
e^{-b(t-s)}S[\br_i(s)]$ for $i=1,2$, the average value of the
difference of both quantities, which is the contribution to
(\ref{deltaC1}) from the interval $(0,t_s)$ will vanish.

To further analyze the behavior of the nonvanishing contribution,
from $(t_s,t)$, we introduce a third trajectory $\br(s)$
satisfying again the Langevin equation (\ref{langevin}) with noise
$\etab$ (independent of $\etab_1$ and $\etab_2$) and endpoint
$\br(t)=\bx$, and consider the time-dependent differences
$\bdri(s) \equiv
\br_i(s)-\br(s)$, for $i=1,2$ (for $0<s<t$). They satisfy the stochastic
equations :
\BE
\frac{d}{ds}\bdri(s)={\bf v}(\br_i(s),s)-{\bf v}(\br(s),s)+
2 \sqrt{\nu} \xibi(s) \ .
\label{diference_full}
\EE
The final values are $\bdr_1(t)=\bdx/2$ and $\bdr_2(t)=-\bdx/2$,
and the new stochastic processes $\xibi(s) \equiv (\etab_i(s) -
\etab(s))/\sqrt{2}$ turn out also to be white noise terms with zero average and
correlation matrix $<\xibi(s)\xibj(s')>=\delta_{ij}{\bf
I}\delta(s-s')$.

As far as the diferences $\bdr$ remain small, the difference in
velocity fields in (\ref{diference_full}) may be linearized so
that
\BE
\frac{d}{ds}\bdri(s)={\bf J}(\br(s))\cdot\bdr_i(s)+ 2\sqrt{\nu}\xibi(s) \ .
\label{differenceJ}
\EE
${\bf J}$ is the Jacobian matrix of the velocity field ${\bf v}$.
Equation (\ref{differenceJ}) can be formally integrated in terms
of the fundamental matrix ${\bf M}(s,t)$, $s \le t$, which is the
solution of
\BE
\frac{d}{ds} {\bf M}(s,t) =
{\bf J}(\br(s))\cdot {\bf M}(s,t)
\label{fundamental}
\EE
with final condition ${\bf M}(t,t)={\bf I}$. The result is:
\BE
\br_i(s)=\br(s)+\bdri(s)=\br(s) \pm {\bf M}(s,t) \cdot \frac{\bdx}{2} +
{\bf G}_i(s,t)\ ,
\label{evolution}
\EE
The positive sign applies to $i=1$, the negative to $i=2$, and
\BE
{\bf G}_i(s,t) \equiv 2\sqrt{\nu} \int_t^s ds' {\bf M}(s,s')
\cdot \xibi(s')\ .
\label{G}
\EE
${\bf G}_i$ is independent of $\bdx$. If the flow ${\bf v}(\bx,t)$ is
chaotic, the matrix ${\bf M}$ will in general produce exponential
growth of the second term in (\ref{evolution}) and of the variance
of ${\bf G}_i$. Thus the linearization leading to
(\ref{differenceJ}) will only be justified for $|s-t|$ small
enough. This may be used to define $t_s$
as the smallest value of $s$ for which linearization is still
valid. From the assumed smallness 
  of $\bdx$, and if $t_s<s$, the second term in
the r.h.s. of (\ref{evolution}) will be small, so that one can
write
\BA
S[\br_i(s)] &\approx& S[\br(s)+{\bf G}_i(s,t)]  \nonumber
\\ &\pm&
\nabla S\left[\br(s)+{\bf G}_i(s,t)\right] \cdot
{\bf M}(s,t)
\cdot
\frac{\bdx}{2} + {\cal O}(\bdx^2)
\label{S}
\EA
where again the two signs refer to the two values of $i$. We are
now in conditions to estimate the contribution to (\ref{deltaC1})
arising from $(t_s,t)$:
\BA
\left<\int_{t_s}^t ds e^{-b(t-s)}\delta S[\br_1(s),\br_2(s)]
\right>_{\etab_1,\etab_2} \approx && \nonumber \\
\left<\int_{t_s}^t ds e^{-b(t-s)} S[\br(s) + {\bf G}_1(s,t)] \right>_{\etab_1}
&& \nonumber
\\ -\left<\int_{t_s}^t ds e^{-b(t-s)} S[\br(s) + {\bf G}_2(s,t)]
\right>_{\etab_2} &&\nonumber
\\ +\left< \int_{t_s}^t ds e^{-b(t-s)} \nabla
S[\br(s)+{\bf G}_1(s,t)]
\cdot
{\bf M}(s,t) \cdot \frac{\bdx}{2} \right>_{\etab_1} &&
\nonumber \\
+ \left<  \int_{t_s}^t ds e^{-b(t-s)} \nabla S[\br(s)+{\bf
G}_2(s,t)]
\cdot
{\bf M}(s,t) \cdot \frac{\bdx}{2} \right>_{\etab_2} && \ .
\label{aveS-S}
\EA
Since $\etab_1$ and $\etab_2$ have exactly the same statistical
properties, the first two averages in the r.h.s. are identical,
and cancel out. The next two averages are also identical and add
up, so that (\ref{deltaC1}), which only receives contributions
from $(t_s,t)$, can be written
\BA
&&\delta\phi(\bx,t;\bdx) =\nonumber \\ &&
\left< \int_{t_s}^t ds e^{-b(t-s)} \nabla S[\br(s)  + {\bf G}_1(s,t)]
\cdot
{\bf M}(s,t) \cdot \bdx \right>_{\etab_1} \ .
\label{deltaSaverage}
\EA
The matrix ${\bf M}(s,t)$ contains the quantitative details on the
exponential separation of trajectories. In the deterministic
dynamics, for which $\bdr(s) \equiv {\bf M}(s,t)\cdot \bdx$, the
time scale of exponential separation is given by the largest
Lyapunov exponent $\lambda$, so that $\bdx(s)$ becomes aligned
with the unit vector along the local unstable direction ${\bf
c}[{\bf r}(s)]$ in a time of order $\lambda^{-1}$, so that
\BE
\bdr(s) \approx {\bf
c}[\br(s)] e^{-\lambda (s-t)} {\bf c^\dagger}(\bx)\cdot\bdx
\label{growth}
\EE
if $|s-t|>\lambda^{-1}$. ${\bf c^\dagger}(\bx)$ is the unit vector
dual to ${\bf c}(\bx)$. 
Both vectors are time dependent for time-dependent flows, but
we do not indicate such dependence for notational simplicity.
In this backward dynamics, the relevant Lyapunov exponent
is the most negative one. But in twodimensional incompressible
flow, it is equal in absolute value to the largest positive one in
the forward dynamics. In (\ref{growth}) we take $\lambda>0$ and
the sign is explicitly written. Analogously, the local expanding
direction in the backwards dynamics is the contracting one in the
forward dynamics. Expression (\ref{growth}) will be introduced in
(\ref{deltaSaverage}). It should be said however that the integral
in (\ref{deltaSaverage}) contains contributions also for
$|t-s|<\lambda^{-1}$. The slow convergence of the Lyapunov
exponent to its asymptotic value \cite{Goldhirsch} 
may introduce corrections, specially if $b$ is large. In
accordance with our aim of neglecting any intermittency
corrections, we approximate the action of ${\bf M}$ by
(\ref{growth}) in all the range of integration.

We now use the mean value theorem to take
out of the integral the temporal average of
$\nabla S \cdot {\bf c}({\bf r}(s))$, which 
we call
$\overline{\Delta S}$. We introduce
in (\ref{deltaSaverage}) the change of variables $u =
e^{\lambda(t-s)}$and write $t_s=t-\tau$ ($\tau>0$). On physical
grounds $\tau \rightarrow
\infty$ if $\bdx$ {\sl and} $\nu\rightarrow 0$, and it is large
but finite for small $\bdx$ and $\nu$. Taking the limit
$t\rightarrow\infty$ and regarding that, in the large time limit,
the statistical properties of the concentration field 
are steady:
\BE
\delta\phi(\bx,\infty;\bdx) \approx  {\bf
c}(\bx)^\dagger
\cdot \bdx
\left<
\frac{\overline{\Delta S}}{\lambda} \int_1^{e^{\lambda\tau}} du
u^{-\frac{b}{\lambda}}\right>_{\etab_1}
\label{integrals}
\EE
Since $\overline{\Delta S}$ is independent of $\bdx$, the scaling
with this last quantity is determined by
\BE
\delta\phi(\bx,\infty;\bdx) =
A {\bf c}(\bx)^\dagger \cdot \bdx \left(  \left<
e^{\tau(\lambda-b)}\right>_\tau - 1
\right)
\label{casifinal}\EE
The only stochastic quantity in (\ref{casifinal}) is $\tau=t-t_s$,
so that the problem has been reduced to the calculation of the
statistics of this quantity. In terms of the characteristic
function of $\tau$, defined as $W(z)\equiv\left< \exp{(-z\tau)}
\right>_\tau$, Eq.~(\ref{casifinal}) reads
\BE
\delta\phi(\bx,\infty;\bdx) =
B   \left[ W(b-\lambda) - 1
\right] \delta x
\label{final}
\EE
where $B=A\cos(\alpha)$, with $\alpha$ the angle of $\bdx$ with
the local expanding direction, and $\delta x =|\bdx|$. Since $t_s$
can be estimated from (\ref{evolution}) as
$|\bdr_{1,2}(t_s)|\approx L$, the calculation of the statistics of
$\tau$ is a classical first passage-time problem, in which one
looks for the time it takes a stochastic process to reach a given
level. There is a vast literature on such problem for stochastic
processes of the form (\ref{evolution}) in which ${\bf M}$ is a
constant matrix or a constant number 
\cite{muchos}.

 In the same spirit as before, we use
Eq.~(\ref{growth}) to approximate (\ref{evolution}) by
\BE
\bdri(s) \approx e^{-\lambda(s-t)}{\bf c}[\br(s)] \left(
\pm {\bf c}(\bx)^\dagger \cdot \frac{\bdx}{2} + 2\sqrt{\nu} F(s,t)
\right)
\label{evolgrowth}
\EE
with
\BE
F(s,t) \equiv \int_t^s ds' {\bf c}[\br(s')]\cdot \xibi(s')
e^{-\lambda(t-s')}
\label{F}
\EE
where we have used (\ref{growth}) inside the integral in (\ref{G}).

If $|t-s| \gg \lambda^{-1}$, $F$ becomes a constant random number,
$F_\infty$. From its definition, it is a Gaussian random number,
of zero average and variance $<(F_\infty)^2>=1/(2\lambda)$. Thus
from (\ref{evolgrowth}) we can write the equation defining $\tau$:
$|\bdri(t_s)| \approx L$, with $t_s=t-\tau$:
\BE
L \approx e^{\lambda\tau}  \left| {\bf c}(\bx)^\dagger
\cdot \frac{\bdx}{2} + \sqrt{\frac{2\nu}{\lambda}} g
\right|
\EE
where $g$ is a random Gaussian number of zero average and unit
variance, from which
\BE
\tau=\frac{1}{\lambda}\ln  {2 L \over
\left| \gamma \delta x + 2 \sqrt{\frac{2\nu}{\lambda}}g \right| }
\label{FPT}
\EE
We have introduced $\gamma=\cos(\alpha)$. In the following we 
denote $L_d= 2 \sqrt{\frac{2\nu}{\lambda}}$, which introduces, as will be 
be seen below, the diffusive length scale.
 The statistics of
$\tau$ may be thus obtained from a change of variables from
the Gaussian statistics of $g$. The characteristic function $W(z)$ can be now
written in terms of (\ref{FPT}) and the Gaussian distribution of
$g$:
\BE
W(z)=\left<e^{-z\tau}\right>=
\frac{(2L)^{-z/\lambda}}{\sqrt{2\pi}}\int_{-\infty}^\infty dg
e^{-\frac{g^2}{2}} \left| \gamma\delta x +
L_d g \right|^{\frac{z}{\lambda}}
\label{W}
\EE

At this point we mention that expression (\ref{final}) with
$W$ given by (\ref{W}) constitutes the main result of this work. 

More explicit expressions can be obtained in the
important cases. If $L_d\ll
\delta x$, then $W(z)\approx (\gamma\delta x/L)^{z/\lambda}$. In
this case,  (\ref{final})
 scales as $\delta x^H$, with
$H=\min(1,b/\lambda)$, in agreement with previous results \cite{prl,pre}.
On the contrary, for small scales, that is,  if
$\delta x \ll
L_d$, then
$W(z)=\pi^{-1/2}(4\nu/L^2\lambda)^{z/\lambda}
\Gamma(\frac{z+\lambda}{2\lambda})$, and $H=1$, thus 
$\delta\phi$
is smooth below a diffusive scale given by $L_d \equiv
2 \sqrt{\frac{2\nu}{\lambda}}$. 
Therefore, we realise of the existence of a crossover length-scale, 
the diffusive scale, given
by $L_d$ which separates a diffusion-controled
 smooth behaviour of $\delta\phi$ 
from an advection-controled with scaling 
 exponent given by $\min(1,b/\lambda)$. This last is
non-smooth when $b<\lambda$.

 We can obtain an approximate
expression simpler than (\ref{W}) by realizing  that the
fluctuations in $\tau$ are much smaller than the average value if
$\delta x$ and $\nu$ are small, so that $<e^{-z\tau}>\approx
e^{-z<\tau>}$, and then substitute the average value $<\tau>$ by
an estimation of it, $\overline{\tau}$, obtained for example by the
condition that the second moment of the trajectory separation
reaches the system size:
\BA
L^2=\left<\bdri(\overline{\tau})^2\right> &\approx&
\frac{e^{2\lambda\overline{\tau}}}{4}\left<\left(\gamma\delta x
+L_d g  \right)^2\right>  \nonumber \\ &=&
\frac{e^{2\lambda\overline{\tau}}}{4}\left(\gamma^2\delta x^2
+L_d^2  \right)
\EA
from which
\BE
\overline{\tau}=\frac{1}{\lambda}
\ln {2L \over \sqrt{\gamma^2\delta x^2+L_d^2}}
\EE
which is an approximation to the mean first passage time $<\tau>$
if $\delta x$ and $\nu/\lambda$ are small. Thus
\BE
W(z)\approx \left( { \sqrt{\gamma^2\delta x^2
+L_d^2}   \over  2L}  \right)^{\frac{z}{\lambda}}
\EE
and
\begin{eqnarray}
\delta \phi(\bx,\infty;\bdx)  \approx C \cos(\alpha )
\left[ \left( \gamma^2\l^2 +l_d^2 \right)^{\frac{b-\lambda}{2\lambda}} -
2^\frac{b-\lambda}{\lambda}
\right]l \ .
\label{clave}
\end{eqnarray}
We have introduced $l\equiv \delta x/L$ and  $l_d \equiv  L_d/L$,
and $C$ is a constant equals to $AL/2^{\frac{b-\lambda}{\lambda}}$. 

The first order structure function, $S_1$, can be obtained by averaging
(\ref{clave}) over the different spatial points $\bx$. Here we realise 
that the point spatial dependence of (\ref{clave}) is only through
the angle $\alpha$ between $\bdx$ and the local expanding
direction. Assuming that these directions are isotropically
distributed, $S_1$ is calculated as:
\begin{eqnarray}
S_1(l) =<|\delta \phi(\bx,\infty;\bdx)|> \nonumber \\
= \frac{C}{2\pi} \int_{0}^{2\ \pi} d\alpha \left| \cos(\alpha)
\left[ \left( \cos(\alpha)^2\l^2 + l_d^2 \right)^{\frac{b-\lambda}{2\lambda}} -
2^\frac{b-\lambda}{\lambda}
\right]l\right| .
\end{eqnarray}

The integral can be performed and we obtain:
\begin{eqnarray}
S_1(l) = \frac{2\ C}{\pi}
  l \Bigg( \left( l^2 + l_d^2 \right)^{\frac{b-\lambda}{2\lambda}}
&  F\left( \frac{1}{2},\frac{\lambda-b}{2\lambda},\frac{3}{2};
\frac{l^2}{l^2 
 +l_d^2} \right)  \nonumber \\
& +2^\frac{b-\lambda}{\lambda} \Bigg),
\label{EST}
\end{eqnarray}
being $F(a,b,c;x)$ the confluent hypergeometric function 
\cite{abramovitz}.
The important thing to be noted in (\ref{EST}) is that $S_1 (l)$ shows
the same scaling behaviour as $\delta \phi$, with a  crossover 
length-scale  given by $l_d =  2\sqrt{\frac{2 \nu}{\lambda}}/L$. 
That is, $S_1(l) \sim l$ for $l < < l_d$ and $S_1(l) \sim l^{\frac{b}{\lambda}}$
when $l >> l_d$ and $b<\lambda$.
Remarkably, 
the diffusive scale, $l_d$,  is independent of the
decaying coeficient, $b$, and therefore its value is the same to the one obtained
for the passive scalar with infinite lifetime, i.e., with $b=0$.

Next section is devoted to check numerically some of the above analytical results.

\section{Numerical results}

In this section we will check the two regimes found for the
scaling of the structure function, and also the expression Eq.
(\ref{EST}). 

Numerically, we proceed by integrating backwards in time Eq.
(\ref{langevin}) with initial conditions 
on a one-dimensional transect of the scalar field. Many
different trajectories are obtained in this way for different
realizations of the noise. Then we integrate Eq. (\ref{FK})
forward in time for each one of these trajectories
($\phi_0 =0$), and finally,
we average over the different trajectories.
 With this procedure we obtain the
field $\phi ({\bf x},t)$, at a fixed time $t$ large enough, on
 a one-dimensional transect without having
to calculate the whole two-dimensional field. Obviously, this
allows us to reach a high resolution for the structure functions
which are then only calculated on a $1d$ transect. For the flow,
${\bf v}=(v_x,v_y)$, we take a simple time-periodic velocity field
defined, in the unit square, $L=1$, and with periodic boundary conditions,
by
\begin{eqnarray}
v_x(x,y,t)& =& -\frac{2U}{T} \Theta \Bigl(\frac{T}{2}-t \bmod T
\Bigr) \cos({2\pi y}) \nonumber \\ v_y(x,y,t)&=& -\frac{2U}{T}
\Theta \Bigl(t \bmod T-\frac{T}{2} \Bigr) \cos({2\pi x})
\label{flow}
\end{eqnarray}
where $\Theta(x)$ is the Heavyside step function, and $T$ is the
periodicity of the flow. In our
simulations $U=1.0$, which produces a flow with a single connected
chaotic region  in the advection dynamics. The value of the
numerically obtained Lyapunov exponent is $\lambda \approx
2.35/T$.  The source term used is $S (x,y) = 0.2 \sin ( 2\pi x)
\sin (2\pi y)$.  We perform our calculations until a final typical time of
$t_f \sim 15T$ where we realise that a final statistical 
stationary concentration field
is reached.

In Fig. (1) we show (in logarithmic scale) the structure functions
against  the length-scale in units of the $l_d$. 
We plot the curves obtained
for different values of the decaying coefficient 
$b=0.5, 0.9, 1.2$, and for a
fixed value of the diffusivity $\nu=5\times 10^{-7}$, and $T=1$.
 For any of the different plots we observe
three different regimes. First, for small scales and up to the
diffusion length scale $l_d$
 ($\ln (l/l_d)=0$)
 the slope of the curve is $1$, showing that the structure function is
smooth in this length-scale interval. 
 Then, for scales larger than $l_d$
and up to typical scales comparable to the system size $L=1$,
appears the other scaling  regime. 
 A straight
line with slope $\frac{b}{\lambda}$ is observed here.
In the figure, we also plot the straight
lines showing this behaviour, and with the number above indicating
their slope.
\vspace{0.5cm}
\begin{figure}
\epsfig{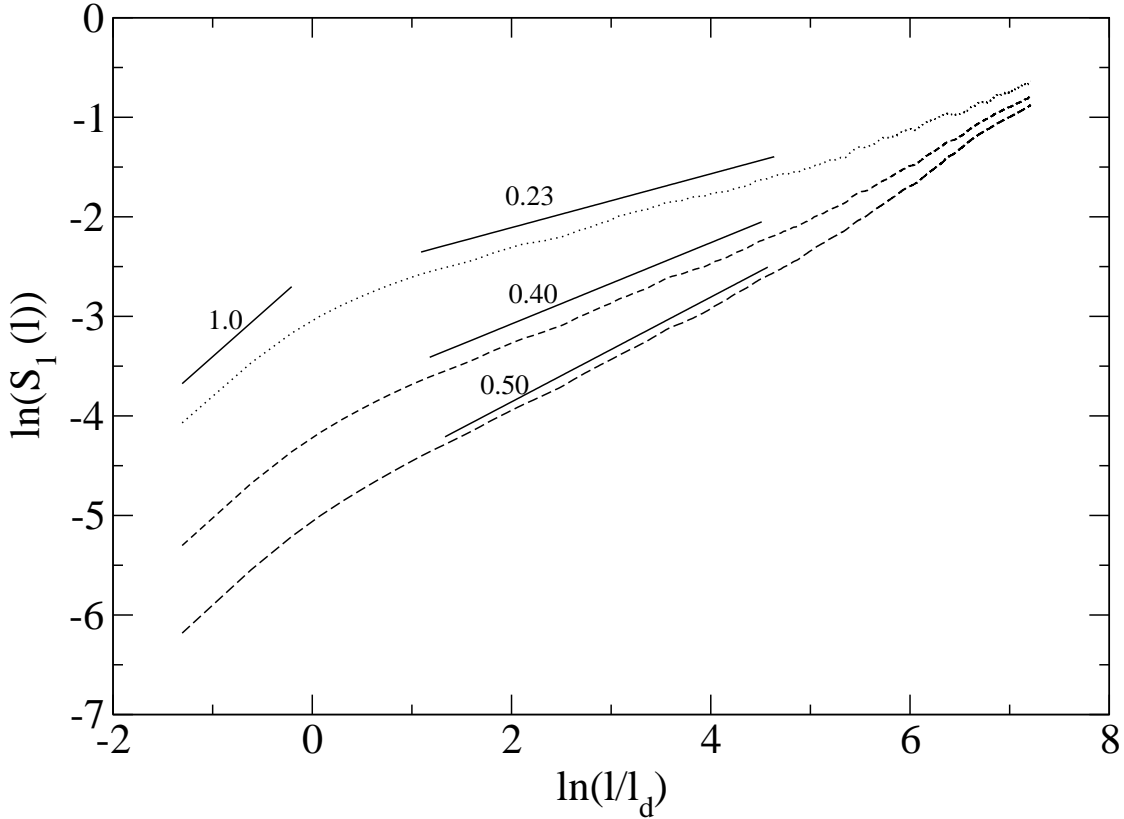}
\caption{This figure plots $\ln (S_1(l))$ against $\ln (l/l_d)$. 
Always $T=1$ and $\nu = 5\times 10^{-7}$.
Here $b=1.2$ for the dotted line,
$b=0.9$ for the dashed line, and $b=0.5$ for the long-dashed line.
The straight line above each curve shows the scaling behaviour,
and the number over the line indicates its slope. The two
regimes are observed: a) smooth (slope
 $1$ for $\ln (l/l_d)< 0$,
 b) scaling behaviour with slope
$b/\lambda$ for $\ln(1) < \ln (l/l_d)<< \ln (L/L_d) = \ln (1/l_d)$. 
}
\label{fig:regimes}
\end{figure}

Next, we proceed to check numerically the validity of Eq.(\ref{EST}).
First, we realize that, on dimensional grounds, the adimensional ratio
$\overline S_1 \equiv S_1(l)/S_1(l=l_d)$ should be a function 
just of the adimensional parameters $l$, $l_d$, and $b/\lambda$. 
Expression (\ref{EST}) fully satisfies this requirement.
This can be used to compare numerically-obtained values of $S_1$ as
a function of $l$ (or $l/l_d$ as in the plot)
 for a variety of values of $\lambda$, $b$, and
$\nu$ with the theoretical prediction. This last
function would be a single curve as long as the
combinations $l_d=2\sqrt{2\nu /\lambda}/L$ and $b/\lambda$ do not change.
Figure (2) shows (for two different values of $(\nu/\lambda, b/\lambda )$)
that the numerical data indeed scale as expected from the theory,
and that the data collapse into the analytic curves is very good,
confirming the accuracy of expression (\ref{EST}).

Finally, in Fig. (3) we perform an analogous plot (in logarithmic scale)
 to 
Fig. (2a)  but for the power spectrum (also normalised
to the diffusion length scale) analytically
calculated by Corrsin
(\ref{corrs}), and  the ones obtained numerically for the same values
of  $b$, $\lambda$ and $\nu$.
 Just a visual inspection confirms our expectation
that, due to the effect of intermittency, it is more 
reliable to compare
with numerical or real data the approximations to
the first order structure function than the approximations to
the power spectrum.
 Also, it is worth to mention that
  to calculate the power
spectrum much more data are needed to obtain a nice statistics
than in the case of the first order structure function.

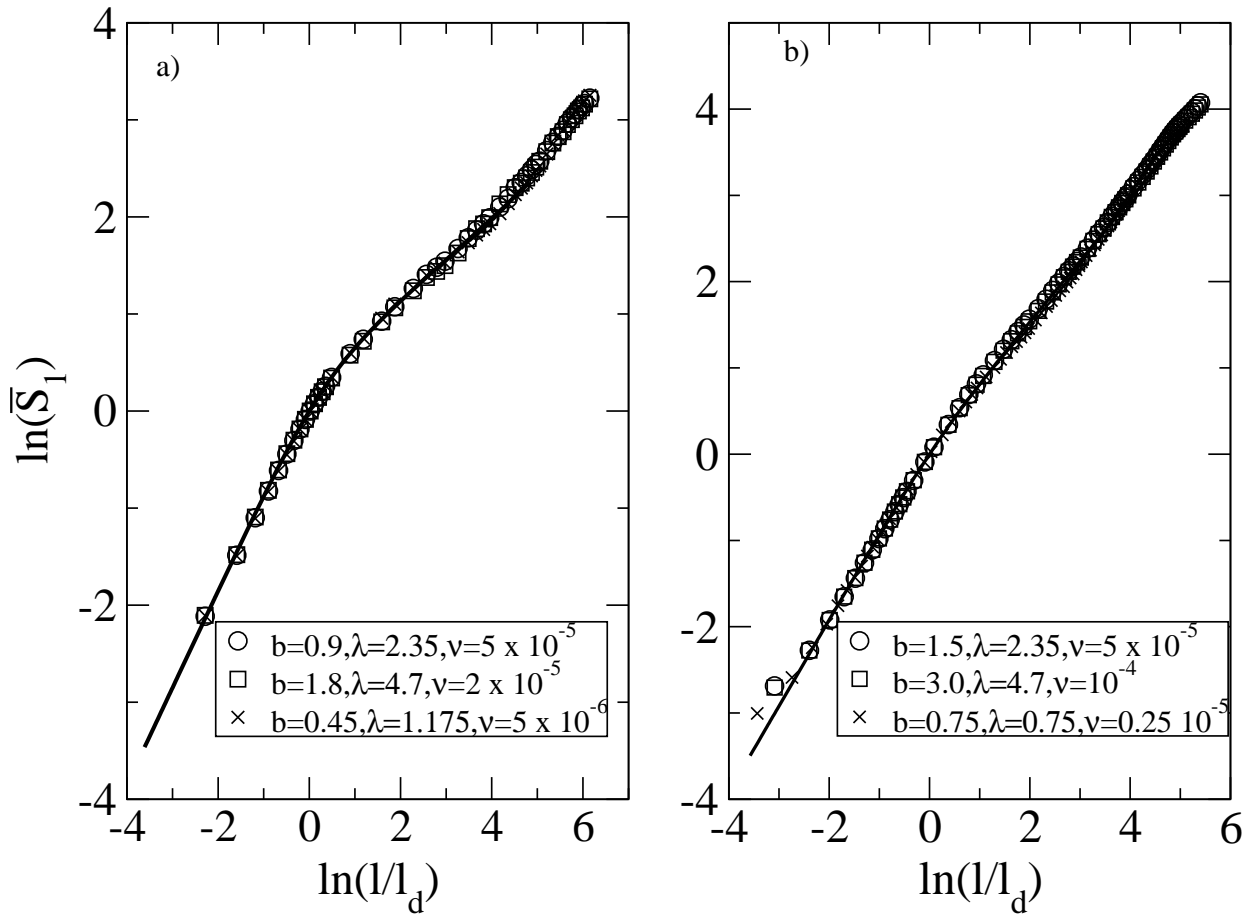
\vspace{1.5cm}
\begin{figure}
\epsfig{file=fig2defi.eps,width=1.0\linewidth }
\caption{Plot of the adimensional  structure function,
$\overline S_1$, as a function of $l/l_d \equiv \delta x/L_d$.
With solid line we plot the theoretical result (Eq.(\ref{EST})), and with
different symbols the different values obtained numerically.
a) is for the  fixed values 
$b/\lambda=0.382$ and $\nu/\lambda=4.2\times 10^{-6}$.
b) the same for
$b/\lambda=0.639$ and $\nu/\lambda=2.1\times 10^{-5}$.
The symbols in the legends indicate the values of $b$, $\lambda$ and
$\nu$ used to obtain the numerical data.
}
\end{figure}

\vspace{1.5cm}
\begin{figure}
\epsfig{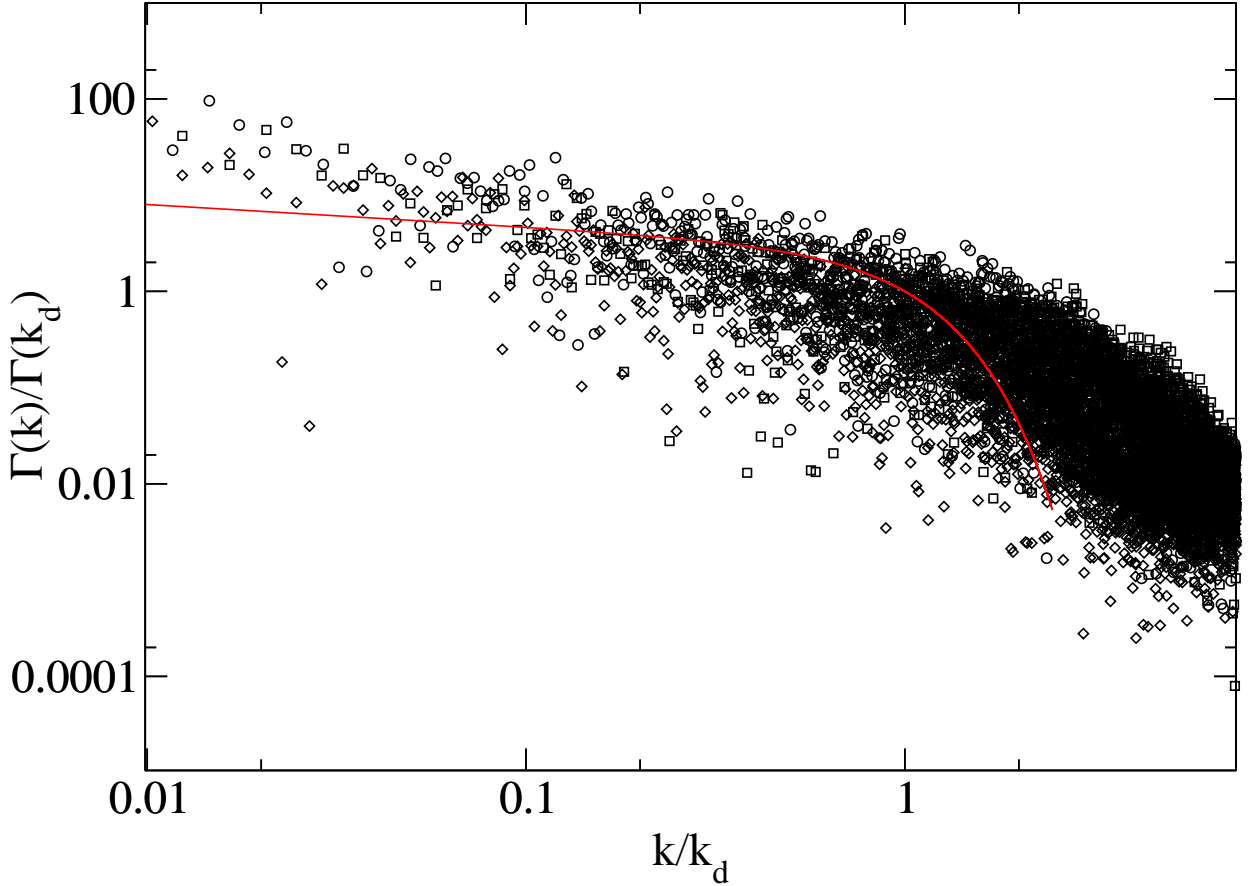}
\caption{Comparison of the power spectrum (solid line)
obtained by Corrsin (\ref{corrs}) and numerical data as in 
Fig. (2a).
}
\end{figure}

\section{SUMMARY}

To summarize, in this paper we have investigated the role of
diffusion in the scaling properties of the first order
structure function of a
passive scalar chaotically advected.
 Based on the Feynman-Kac
representation of the advection-reaction-diffusion equation, we
have obtained an analytical expression for the
 the shape of the first order
structure function for all scales smaller than the typical system
size, and observed the existence of a crossover between a smooth
regime and a scaling regime with slope depending on the rate of
the decaying coefficient and the maximum Lyapunov exponent of the
flow. The length-scale for the crossover is the typical
dissipative length scale, given by
$L_d=2\ \sqrt{\frac{2\nu}{\lambda}}$, which  is independent of
the decaying coeficient and, therefore, is the same as for the
passive scalar with infinite lifetime.
Moreover, our results confirm that the scaling exponent of $S_1$ are
not modified by the molecular diffusion.
We have shown that our approximation for the first order 
structure function is more reliable than the one for the power
spectrum obtained under the same disregard of intermittency
corrections. This could be relevant when explaining or
modeling numerical or experimental data.
 We have also provided numerical support to
our analytical findings.
These are in excellent agreement. 
 Our numerical algorithm uses   the
Feynman-Kac representation, and allows for a very fine scale
resolution, as we just need to calculate the scalar field in
one-dimensional sections of the whole surface. We believe that
this numerical algorithm  can be very useful for other kind of
advection-reaction-diffusion problems.


\section{Acknowledgments}

We acknowledge 
Zolt\'an Neufeld for his contributions to an early version of
the Paper, and  
Angelo Vulpiani for suggesting us the use of the
Feynmann-Kac formula. C.L. acknowledges financial
support from the Spanish MECD.
E.H-G acknowledges support from MCyT (Spain) projects BFM2000-1108
(CONOCE) and REN2001-0802-C02-01/MAR (IMAGEN).


\end{document}